\input psfig.sty

\ifx\mnmacrosloaded\undefined \input mn\fi

\pageoffset{-2.5pc}{0pc}

\loadboldmathnames



\pagerange{000--000}    
\pubyear{1998}
\volume{000}

\begintopmatter  

\title{Unambiguous quasar microlensing}

\author{Mark A. Walker}

\affiliation{Special Research Centre for Theoretical Astrophysics, School of Physics,
University of Sydney, NSW 2006, Australia}

\shortauthor{Walker}
\shorttitle{quasar microlensing}

\acceptedline{Accepted \ \ \ \ \ \ \ \ \ \ \ \ \ \ \ \ \ \ \ \ \ \ . Received}

\abstract{Microlensing studies of quasars can reveal dark matter lumps
over a broad mass spectrum; we highlight the importance of monitoring
quasars which are seen through the halos of low-redshift galaxies. For
these configurations microlensing by planetary-mass objects will manifest
itself as isolated events which are only weakly chromatic. Statistical comparison
of the observed optical depths with their theoretical counterparts provides
a strong test for a microlensing origin of such events.
If microlensing is detected, the light-curves can reveal not only the
characteristic microlens masses, and their corresponding contribution to
dark halos, but also how compact the individual objects are. In this way we
can decisively test the possibility that the dark matter associated with
galaxies is composed principally of planetary-mass gas clouds.}


\maketitle  

\section{Introduction}
Flux monitoring of quasars has provided evidence that the dark matter
might be composed of objects of planetary mass, ${\rm M\la 10^{-3}
\,M_\odot}$. This evidence comes from two types of observations:
optical variability that may be due to distant gravitational microlenses
(Irwin et al. 1989; Hawkins 1993; Schild 1996); and radio monitoring data
(Fiedler et al. 1987), which show ``Extreme Scattering Events'' (ESEs),
can be sensibly interpreted as plasma lensing by cool clouds in the Galactic
halo (Walker \& Wardle 1998). However, these data can be interpreted
in other ways (Wambsganss, Paczy\'nski \& Schneider 1990; Baganoff \&
Malkan 1995; Schmidt \& Wambsganss 1998; Romani, Blandford and Cordes 1987),
and the idea that dark matter takes the form of planetary mass lumps
is currently just an interesting suggestion. What is needed
now is to move away from suggestive evidence, which has served its
purpose in drawing attention to the proposed picture, and towards some
decisive observational tests. Such tests have previously been contemplated
(Press \& Gunn 1973; Gott 1981; Canizares 1982; Vietri \& Ostriker 1983;
Paczy\'nski 1986), with firm negative results for some mass ranges
(Press \& Gunn 1973; Dalcanton et al. 1994; Carr 1994; Alcock et al. 1998).
However, all of these tests admit the possibility of a
substantial quantity of dark matter residing in planetary-mass
gas-clouds. In view of the quasar monitoring data, observations designed
specifically  to investigate this particular mass range would be worthwhile.
In this paper we demonstrate that
clean experimental tests are, in fact, quite straightforward to arrange
in respect of quasar microlensing. The most important consideration
is selection of the sample of sources to be monitored; as described
in \S2 and \S3, these should be apparently close to low redshift galaxies.
In \S4 we discuss the potential of such data for distinguishing
between low-mass gas clouds and more compact microlenses.

\section{Microlensing at low optical depth}
The large-scale distribution of dark matter within a galaxy can
be approximated by an isothermal sphere, for which
the surface density as a function of radius, $r$, is (Gott 1981)
$$
\Sigma(r)={{\sigma^2}\over{2Gr}},\hfill\stepeq
$$
where $\sigma$ is the line-of-sight velocity dispersion, and we
have neglected the possibility of a core in the density distribution.
Suppose now that this surface density is {\it entirely\/} in compact lumps
of material (the meaning of ``compact'' will be addressed in \S3),
then it follows that the optical depth to gravitational microlensing
by these lumps is just
$$
\tau={{\Sigma(r)}\over{\Sigma_c}}=2\pi{{\sigma^2}\over{c^2}}
{1\over\chi},\hfill\stepeq
$$
where $\Sigma_c\simeq{{c^2}/{4\pi GD_d}}$
is the critical surface density for multiple imaging, and
$\chi\equiv r/D_d$, for a galaxy at distance $D_d$. Here we
have assumed that we are observing a distant quasar behind a
low-redshift galaxy. If, for simplicity, we suppose that
all large galaxies can be approximated as having $\sigma\simeq150\;
{\rm km\,s^{-1}}$, then we arrive at the convenient formulation
$$
\tau\sim{1\over{3\chi}},\hfill\stepeq
$$
where $\chi$ is now expressed in arcseconds. From this we can
see immediately that any quasar which lies within, say, an arcminute
of a bright galaxy stands a modest chance of being microlensed, providing
only that the dark matter is in compact form.

A corollary of the above is that any quasar which is aligned within
about an arcsecond of the centre of an intervening galaxy has near unit
probability of being microlensed at any given time. So is such an
alignment the most favourable place to investigate microlensing? No.
This situation is favourable in only two respects: first there will
very likely be macrolensing -- i.e. multiple imaging of the quasar on the
scale of arcseconds -- which, by photometric monitoring of the individual
macro-images, permits microlensing to be distinguished from variations
which are intrinsic to the quasar. Secondly the large optical
depth means that there will be essentially continuous microlensing
variations. Unfortunately these benefits are offset by two substantial
disadvantages: first the central arcsecond of any large galaxy is composed
predominantly of stars, not dark matter; and secondly the network of
caustics which occurs at high optical depths (see, for example,
Schneider, Ehlers \& Falco 1992) leads to complex light-curves which are
difficult to interpret. (Note that it is not easy even to measure,
accurately, the brightness of the individual macro-images, as they are
blended with the core of the lensing galaxy, and with each other, in
ground-based observations). Indeed these problems are evident in the literature
on the gravitationally lensed quasar 2237+0305, which is seen through
the central bulge of a low-redshift galaxy. Rapid variations in the flux of
one of the macro-images were initially interpreted (Irwin et al 1989)
in terms of microlensing by a low-mass object, but this interpretation
was later challenged (Wambsganss, Paczy\'nski \& Schneider 1990)
and the data re-interpreted in terms of microlensing by stars.  Subsequently
it has been emphasised that the light-curves for this system {\it do\/} admit a
population of low-mass lenses (Refsdal \& Stabel 1993), leaving the
whole question quite open.

The difficulty of interpreting light-curves which arise from a dense
network of caustics argues for a shift in emphasis towards monitoring
of systems where the optical depth is small.
In this regime we may observe individual microlensing events, superimposed
on a more-or-less steady baseline (distinguishing microlensing from other
forms of variability is addressed in \S5). This is a great advantage in that
the observed event time-scales can then be related more-or-less directly
to mass scales. The lower event-rate associated with a small optical
depth is the principal disadvantage of this regime, but this can be offset
by studying a larger number of targets in order to accumulate good
statistics. 
\beginfigure{1}
\psfig{figure=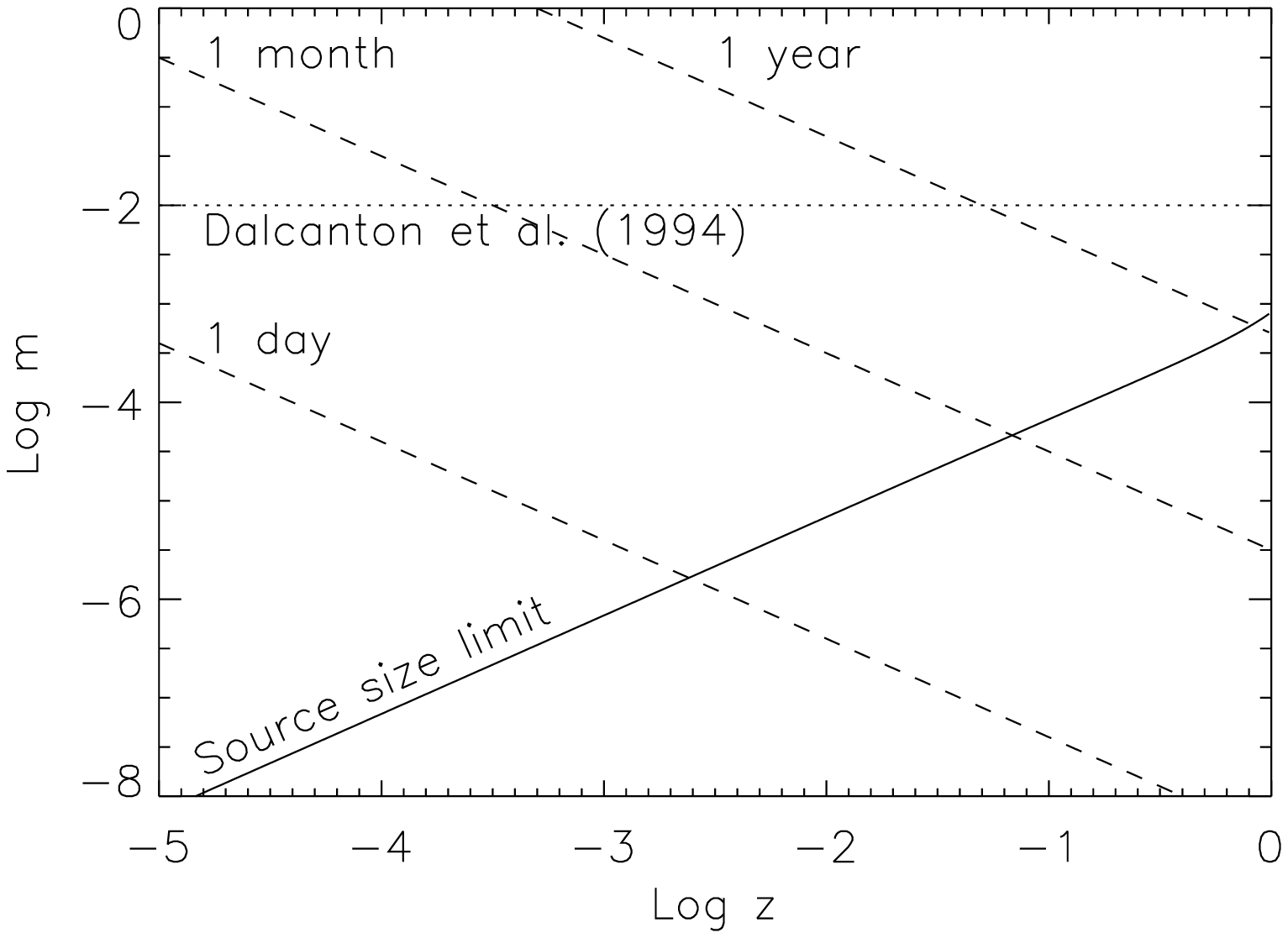,width=8cm}
\caption{{\bf Figure 1.} Detectability of gravitational microlensing
by compact masses ($m\equiv M/{\rm M_\odot}$) as a function of lens
redshift, $z$. In order for a microlens to introduce significant magnification,
it must be more massive than indicated by the ``source-size limit''.
The adopted source size is $10^{15}$~cm, located at $z_Q=2$, and 
an empty ($\Omega=0$) universe is assumed.
Approximate time-scales for microlensing events are shown by the
dashed lines (assuming a transverse speed of $600\;{\rm km\,s^{-1}}$).
The upper bound of Dalcanton et al. (1994) is also plotted: this corresponds
to the largest permitted mass, if galaxy halos contribute $\Omega_g\sim0.1$,
are entirely composed of microlenses, and $\Omega<0.6$.} 
\endfigure

\section{Microlensing at low redshift}
So far we have given no reason to prefer galaxies in any particular
redshift range. In constructing a sample of quasars seen through
galaxy halos we would find that most of the cases involve
distant galaxies, simply because there is a greater surface density
of distant galaxies than nearby ones. Unfortunately these examples
are less useful for investigating microlensing by low-mass objects;
the reason is that at large distances the angular size of the (Einstein
ring of the) lens becomes smaller than the angular size of the quasar, and
so the apparent changes in quasar flux
are small. If this happens we lose not only signal-to-noise ratio
but also our ability to predict light-curves, because our current understanding
of the emission from quasars is so poor that the point-source approximation
is the only one which enables confidence. (Of course, one can use
lensing phenomena to investigate source structure, but that is not
our concern here.) This is especially important because a resolved
source may exhibit substantial differences between the light-curves
seen at different frequencies, whereas microlensing events of a point-like
source by a point-like lens are achromatic and this feature aids the
interpretation of any observed variability.

Figure 1 illustrates the limits imposed by source size on the detectability
of microlenses, of various masses, as a function of lens redshift. (We note
here that the mass limits quoted by Walker \& Ireland [1995] are too pessimistic
-- they appear to have been derived from a comparison between the linear
dimensions, rather than the angular dimensions, of lens and source -- thanks
to Steve Warren for drawing attention to this.) To be definite we take a source
of radius $10^{15}\;{\rm cm}$ (c.f. Wambsganss, Paczy\'nski \& Schneider 1990)
at $z_Q=2$; we adopt a Hubble constant of $75\;{\rm km\,s^{-1}\,Mpc^{-1}}$, and
angular-diameter/redshift relations appropriate to an $\Omega=0$
universe. From figure 1 it is evident that for microlenses at $z_d\sim1$
there will be relatively little sensitivity to the mass range
$M\la10^{-3}{\rm M_\odot}$. This is just the mass range of interest and so
it is critical to select lines-of-sight which intersect {\it low-redshift\/} galaxies.
For example, at $z_d\sim10^{-2}$ we have sensitivity to microlenses of
$M\ga10^{-5}{\rm M_\odot}$, with the upper end of the mass range being fixed by
the duration of the monitoring experiment. We note that if galaxies contribute
$\Omega_g\sim0.1$ to the cosmological density parameter, $\Omega$, and $\Omega<0.6$,
then microlenses which are more massive than $10^{-2}{\rm M_\odot}$
cannot dominate their halos (Dalcanton et al. 1994) --- a result
which follows from analysis of the equivalent widths of quasar emission lines.
This bound is plotted in figure 1. Approximate
microlensing event time-scales are also plotted in figure 1, and from
these we see an added advantage of low redshift galaxies, namely that the
time-scales are well matched to an observing program. By contrast events
involving $10^{-3}{\rm M_\odot}$ microlenses take years at $z_d\sim1$. 

The considerations we have given also apply to
the lowest redshift halo, namely the Galactic halo,
which has a very small optical depth, $\tau<10^{-6}$.
Paczy\'nski (1986) suggested that its compact constituents could be revealed
by their microlensing influence on the flux from LMC stars. Indeed microlensing
by stellar-mass objects has now been detected in this way (Alcock et al.
1997). Precisely because the observed lenses are stellar mass, however, there
remains a concern that these signals are due to the known Galactic/Magellanic
stellar populations (Sahu 1994), or to tidal debris (Zhao 1998), and are
unrepresentative of the Galactic halo as a whole. This notion is reinforced
by the Dalcanton et al. (1994) constraints, mentioned earlier.
Significantly, no signal has been seen from planetary-mass objects
(Alcock et al. 1998), which calls into question the microlensing interpretation
of quasar variability. One possible resolution of this apparent conflict is that
the low-mass microlenses suggested by Irwin et al. (1989), Hawkins (1993,1995)
and Schild (1996) are not actually dense enough to be strong microlenses in the
context of the Galactic halo, implying that their characteristic surface density
lies in the range
$$
0.1\quad \la \quad\Sigma({\rm g\,cm^{-2}}) \quad\la\quad 10^4.\hfill\stepeq
$$
This range includes the estimated mean surface density of the individual gas clouds
($\sim10^2\;{\rm g\;cm^{-2}}$) in the model of Walker \& Wardle (1998),
but excludes black holes and planets. It is worth noting that {\it all\/}
baryonic, Galactic dark matter candidates are required to have a characteristic
surface density $\Sigma\ga3\;{\rm g\,cm^{-2}}$, in order for them not to have
collided with each other within the age of the Universe (Gerhard \& Silk 1996).
This implies that all baryonic dark matter candidates associated with galaxies
must be strong gravitational lenses by $z_d\sim0.03$.

An important point has recently been made by Draine (1998): dense gas clouds
can act as strong lenses purely on account of the refractive index of the
gas itself. Draine further notes that the optical light curves for {\it gas\/}
microlensing events are very similar to those for {\it gravitational\/} microlensing
(see also Henriksen and Widrow 1995), raising the
startling possibility that some of the observed microlensing events
might actually be due to gas lensing! At present predictions
concerning gas lensing are limited primarily by our ignorance of the likely
run of density within the putative clouds. But given any specific density
distribution we can incorporate the refractive index of the gas in
our calculation of lensing behaviour; this is the approach we shall take.
\beginfigure{2}
\psfig{figure=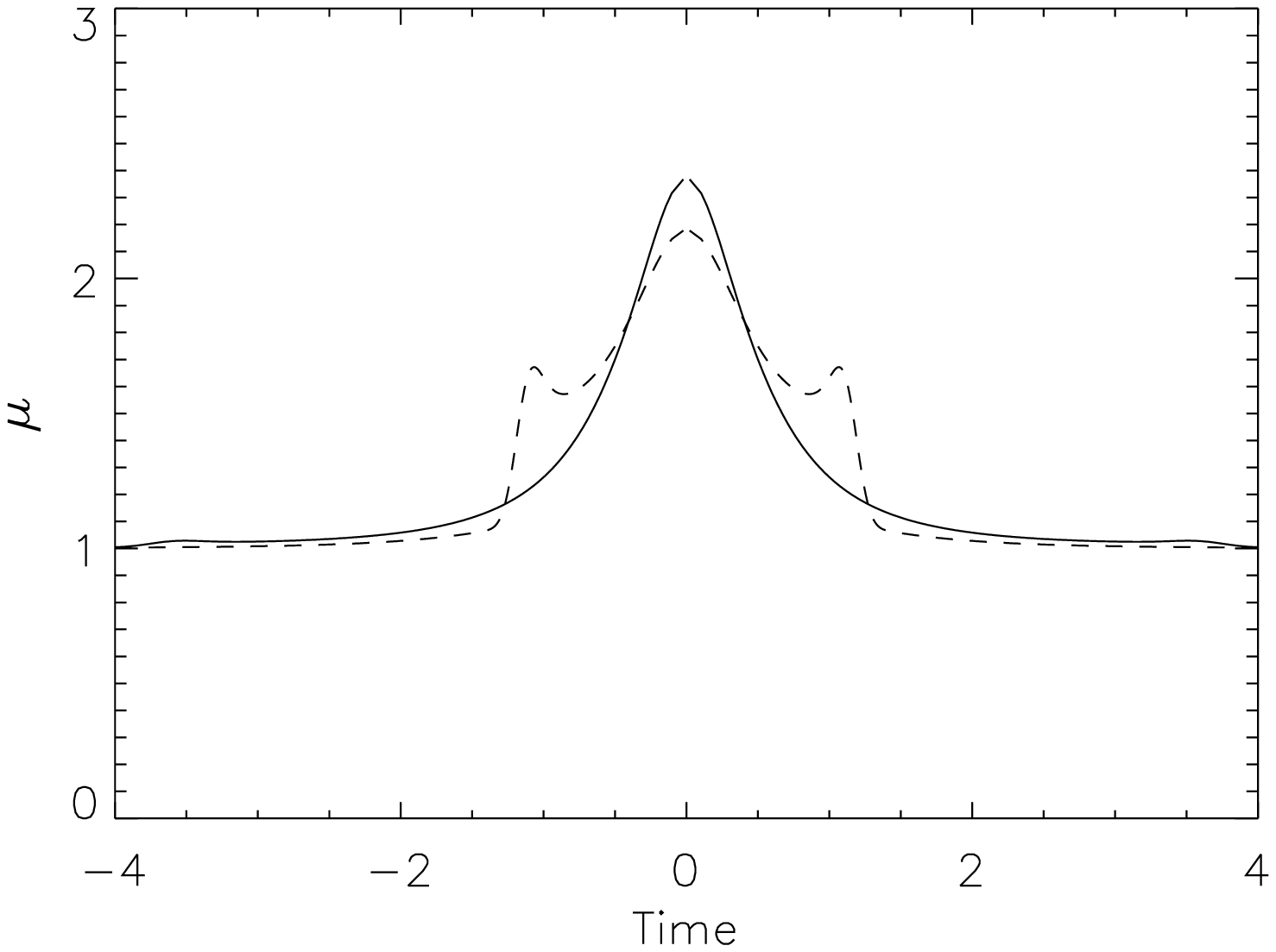,width=8cm}
\caption{{\bf Figure 2.} Model light-curves for lensing of a $z_Q=2$
quasar by a $3\times10^{-4}\,{\rm M_\odot}$ gas cloud, of $\Sigma_0=100\;{\rm
g\,cm^{-2}}$, at $z_d=0.002$ (dashed line) and 0.01 (solid line). These
correspond to $\kappa_0=2.3, 12$, respectively, while $\kappa_0^\prime$
(describing refraction by the gas itself) is, in the optical band, roughly 30\%
larger than $\kappa_0$ in each case. 
The impact parameter for each event is taken to be 0.5 Einstein
ring radii, and time is given in units of the crossing-time for
one Einstein ring radius. The same source model is adopted as for
figure 1.} 
\endfigure

\section{Microlensing by gas clouds}
In the previous sections we concentrated on the means by which
one can best test the picture that dark matter takes the form of planetary
mass lumps, with little regard for the specific nature of these lumps.
The observations which we advocate can, however, tell us
more than just the mass of any microlens: they also give us information
on the surface density distribution of the individual lenses. At a crude
level this is already obvious from our discussion in \S3.
On a more subtle level
there are diagnostic features present in the light-curves even when
the clouds are securely in the strong lensing regime (see also
Henriksen \& Widrow 1995).
To demonstrate this we take the example of a Gaussian surface
density profile for each microlens
$$
\Sigma(r)=\Sigma_0\,\exp(-r^2/2\sigma^2).\hfill\stepeq
$$
The corresponding
mass is then $M=2\pi\sigma^2\Sigma_0$. If we express all angles
in units of the Einstein ring radius for this mass then we arrive at
the lens equation which gives the image locations ($\theta$)
implicitly in terms of the source location ($\beta$):
$$
\beta = \theta\left[1-\kappa_0^\prime\exp(-\kappa_0\theta^2)\right] - 
{1\over\theta}\left[1-\exp(-\kappa_0\theta^2)\right]. \hfill\stepeq
$$
Here we have written $\kappa_0\equiv\Sigma_0/\Sigma_c$ for the
central surface density of the cloud in units of the critical surface
density for multiple gravitational imaging; and similarly
$\kappa_0^\prime\equiv\Sigma_0/\Sigma_c^\prime$ where, following
Draine (1998), and making use of his quantity $\alpha$,
$\Sigma_c^\prime\simeq\sigma^2/\alpha D_d=M/2\pi\Sigma_0\alpha D_d$.
In the optical band, $\alpha\simeq1.2\;{\rm cm^3\,g^{-1}}$, not strongly
dependent on frequency (Draine 1998). In equation 6, then, the term in
$\kappa_0^\prime$ describes refraction by the gas.

\beginfigure{3}
\psfig{figure=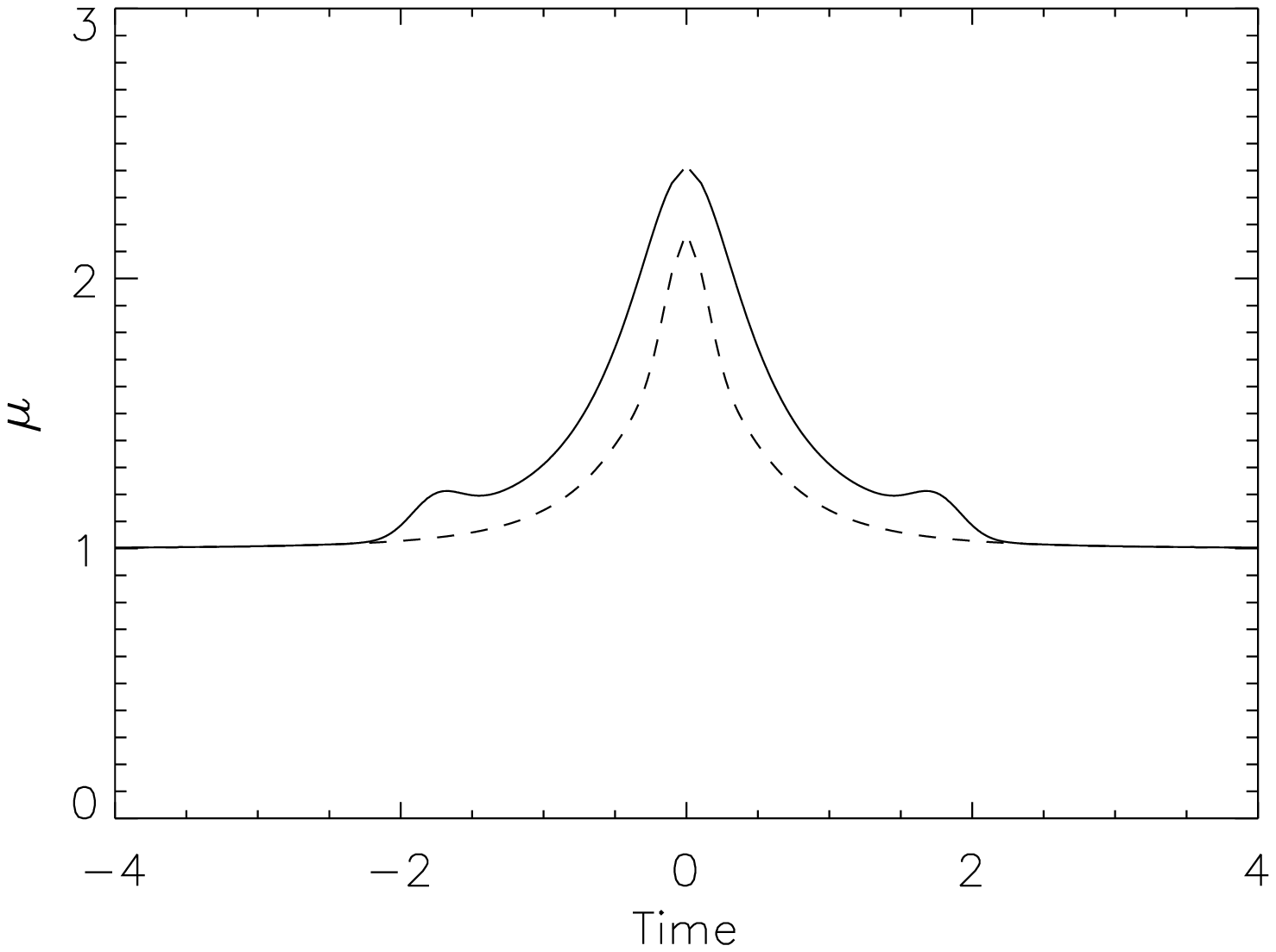,width=8cm}
\caption{{\bf Figure 3.} As figure 2, but with refraction by the gas neglected
($\kappa_0^\prime=0$), so that lensing is entirely due to the gravitational field.}
\endfigure
There are two simple analytic limits of equation 6: for $\kappa_0\theta^2
\gg1$ we recover the Schwarzschild lens mapping $\beta\simeq\theta-1/\theta,$
appropriate to the point-mass lens approximation; while at the other
extreme ($\kappa_0\theta^2\ll1$) we have $\beta\simeq\theta(1-\kappa_0-\kappa_0^\prime)$
as, indeed, we might expect (see Schneider, Ehlers \& Falco, 1992,
for discussion of these cases). The individual image magnifications are
determined from $\mu=(\theta/\beta)\partial\theta/\partial\beta$,
and so the light-curves corresponding to these lenses will evidently
be very different. Of more interest, though, is the general
case for which we require the exact mapping
(equation 6). In figure 2 we show theoretical light-curves for clouds
of mass $3\times10^{-4}{\rm M_\odot}$, and central surface density
$\Sigma_0=100\;{\rm g\,cm^{-2}}$, at $z_d=0.002,0.01$. These curves
pertain to the optical band, where refraction by the gas itself contributes
substantially (Draine 1998) --- for our particular model $\kappa_0^\prime
\simeq4\kappa_0/3$.
It is reasonable to anticipate that target quasars could be monitored
with a standard error of 0.01~magnitudes, so the differences between
these two light-curves are easily measurable. The more distant of the two
examples is not quite distinguishable from a truly point-like lens.
An important qualitative feature of each curve, which is not present for
the Schwarzschild lens, is the existence
of a fold caustic at $\theta\simeq1/\sqrt{\kappa_0+\kappa_0^\prime}$.
This caustic introduces a thin annulus of high magnification which,
by virtue of its small angular extent, is expected to be chromatic even
if the principal peak in the light-curve is not; this caustic is evident
in figure 2 as the subsidiary peaks at $t\simeq\pm1.2$ for the lower redshift
lens. For the more distant lens the caustic crossing occurs at $t\simeq\pm3.7$,
but the high magnification region is so thin (in comparison with the
source dimension) that there is no peak in the light-curve at these locations.
For reference we show in figure 3 light-curves for our model clouds
at wavelengths where there is negligible refraction by the gas itself
(i.e. $\kappa_0^\prime\ll\kappa_0$). In figure 3 both light curves are
readily distinguished from microlensing by a point-mass lens. Relative
to figure 2, where refraction by the gas is non-negligible, the principal
difference is that the caustic ring shrinks in radius, because of the
smaller central ``beam convergence'' (sum of $\kappa_0$ and $\kappa_0^\prime$),
and becomes broader. For the more distant lens (solid line), the increased
width of the annulus of high magnification renders the caustic more
visible in figure 3 than figure 2. For the lower redshift lens, however,
the source only grazes the caustic, rather than crossing it, and this
leads to a single central peak in the light-curve.

\beginfigure{4}
\psfig{figure=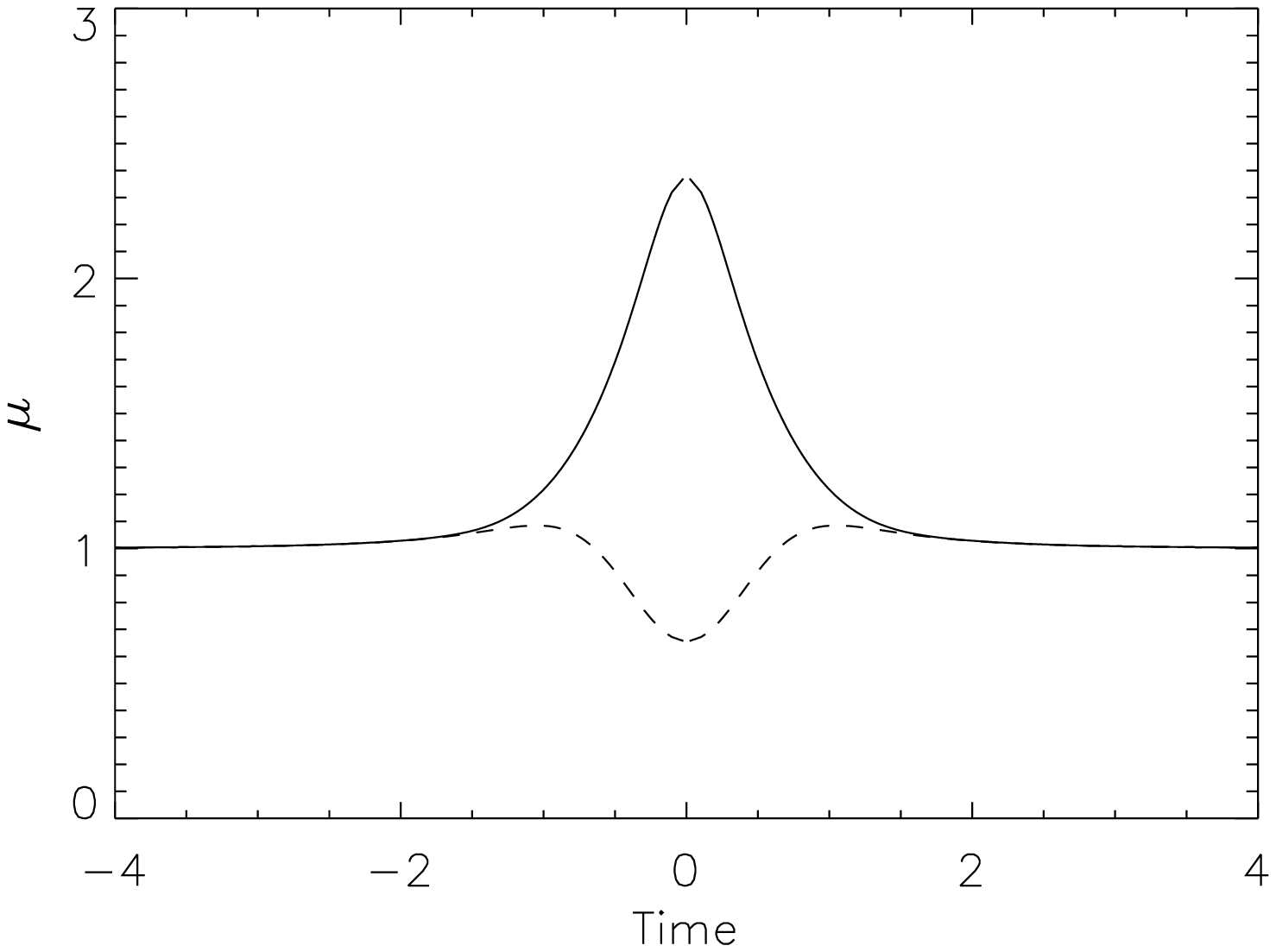,width=8cm}
\caption{{\bf Figure 4.} As figure 2, but with $\kappa_0^\prime=0$, and
adopting an opacity of $0.4\;{\rm cm^2\,g^{-1}}$ --- these values are appropriate
to lensing in the hard X-ray band, where Thomson scattering dominates the
opacity of the gas.}
\endfigure
A final point to make is that where light passes through a gas cloud some
absorption may occur. This is expected to be a small effect in the optical
band (else the putative clouds should have already been discovered in this way),
but the extinction is certainly large in the far UV and throughout the X-ray
region. X-ray light-curves are therefore expected to appear quite different
to their optical counterparts, in cases where the lens is not point-like. 
To demonstrate this we have computed X-ray light-curves for our model
clouds. At energies of several keV to several hundred keV the extinction
is principally due to electron scattering, so across this broad range each
cloud presents an optical depth $\tau\simeq40\,\exp(-r^2/2\sigma^2)=
40\,\exp(-\kappa_0\theta^2)$. Image locations are as given by equation 6,
with $\kappa_0^\prime=0$ (refraction by the gas is negligible in the X-ray
band), and each image is attenuated by a factor $\exp(-\tau)$.
The resulting light-curves are shown in figure 4; these curves may be
compared directly with those of figure 2, which involve the same lensing
geometry and differ only in observing wavelength. Interestingly, while the
more distant lens has a light-curve which qualitatively still resembles
microlensing by a point mass, the nearby lens ($z_d=0.002$, dashed curve)
manifests an extinction event. Now both lenses have the same physical optical
depth profile, with a central optical depth of 40, so this difference is entirely
a consequence of the different lensing geometries. More specifically: during both
events magnified images of the source lie close to the Einstein ring of the
lens, i.e. at $\theta\sim1$ in each case; this corresponds to an
optical depth of $40\,\exp(-\kappa_0)$, which is $\simeq4\times10^{-4}$
for the more distant lens, but $\simeq4$ for the closer one, leading
to substantial attenuation of the lensed images in the latter case.
In other words, for sufficiently distant clouds the strongly magnified
images are located at large physical separations from the cloud,
and the lens can be regarded as effectively point-like.

\section{Discussion}
The main barrier to the investigations we advocate is not so much the
actual photometric monitoring, which is routine, but the identification of
suitable targets. One approach which has previously been suggested
(Walker \& Ireland 1995; Tadros, Warren \& Hewett 1998) is
to monitor  quasars lying behind rich clusters of galaxies at
low redshift, but this approach is really only feasible for cameras
which have exceptionally large fields of view.
An alternative is to construct a very large sample of quasars, and then
select out the small fraction which are viewed through halos
of foreground galaxies, by cross-correlating with a galaxy catalogue.

We can estimate the microlensing optical depth which is contributed
by galaxies within redshift $z_d$ ($z_d<1$) from $\tau(z_d)\sim
\Omega_g\, z_d^2$ (c.f. Press \& Gunn 1973), where $\Omega_g$ is the
average mass per unit volume in galaxies, expressed in units of the
critical density, and we have assumed that galaxies are
composed predominantly of microlenses. Taking $\Omega_g\sim0.1$
it follows that we need a sample of $N_Q\sim10^5$ quasars in
order to amass a combined optical depth in excess of unity from
galaxies within $z_d\simeq10^{-2}$; no such sample exists. However,
the dependence on redshift is quadratic, and within $z_d\simeq0.02$
we need only $N_Q\sim3\times10^4$ sources; so with the largest available
quasar survey (Boyle et al. 1998) we expect a combined optical
depth of $\tau(z_d=0.02)\sim1$. Of course the bulk of the
quasars in any survey make a trivial contribution to
this estimate, because they do not lie behind galaxy halos.
If, for example, we suppose that each galaxy halo extends to a
radius of 50~kpc, and we take the space density of galaxies to be
$4\times10^{-3}\;{\rm Mpc^{-3}}$, then only 80 quasars contribute.
Thus by selecting out the close angular coincidences between
quasars and galaxies at low-redshift, it becomes feasible to monitor
a sub-sample in which there is always a microlensing event
in progress. At $z_d\la0.02$ the influence of the quasar dimensions
on the observed light-curves should be small, provided the microlenses
have masses $M\ga3\times10^{-5}\,{\rm M_\odot}$ (see figure 1).  

It is worth setting out the criteria by which microlensing can
be distinguished from other causes of variability. Most importantly, for
the proposed experimental conditions we expect only weakly chromatic
light-curves for the optical continuum. If the lenses are sufficiently
point-like then similar light curves are expected for the X-ray band,
assuming that the X-ray source is comparable in size to the optical
source. (But non point-like lenses introduce attenuation which may
lead to X-ray extinction events --- compare figs. 4 and 2.) However,
one does not expect any associated changes in radio flux,
because the emission region in this case is too large to be significantly
affected. The same holds true for the optical emission lines, which
are believed to arise from a region of much larger dimensions than the
optical continuum. Note that for broad-band optical photometry this means 
there will always be a non-varying component required when fitting
to theoretical light-curves. In this case one needs to subtract the steady,
emission-line flux in each band prior to testing for achromaticity. For a
single lens, and no strong external shear, one also expects the light-curves
to be time-symmetric. These criteria can be employed for individual events.
In addition, a strong test for microlensing becomes possible when a sample
of candidate events is available: correlation of measured optical depth with
the theoretical estimate. This correlation is expected to be good because the
connection between theory (e.g. \S2) and experiment is very close. This,
coupled with the fact that intrinsic variations should have absolutely nothing
to do with foreground objects -- i.e. zero correlation predicted for intrinsic
variations -- makes for a very powerful test indeed.

\section{Conclusions}
It is desirable to initiate photometric monitoring of quasars seen through
the outer halos of low-redshift galaxies. For this type of configuration,
planetary-mass lumps of dark matter introduce discrete microlensing
events which are unlikely to be confused with intrinsic outbursts. A
strong test of any putative microlensing is available in the ensemble properties
of the quasar sample: the measured optical depth should correlate well
with the theoretical value. An observed correlation of this type would eliminate
the possibility of events being intrinsic to the sources, while non-detection
could, for example, eliminate {\it all\/} Jovian-mass dark matter candidates
--- a conclusion which cannot be reached on the basis of LMC microlensing
observations. If microlensing is indeed detected, the observed light-curves
have the potential to differentiate between point-like lenses (black holes/planets)
and gas clouds.


\section*{Acknowledgments}
Thanks to Mark Wardle, Brian Boyle and Ken Freeman for their thoughts on
the various issues herein.

\bye